\documentclass{aastex631}
\usepackage{CJK}
\usepackage{amsfonts}
\usepackage{amsmath}
\usepackage{mathrsfs}
\usepackage{epsfig}
\usepackage{subfigure}

\newcommand\TESS{{\it TESS }}

\def\cd{\,\mathrm{c}\ \mathrm{d}^{-1}}

\def\raa{Res.\ Astron.\ Astrophys.\ }

\shorttitle{Pulsation Analysis of HADS with TESS}
\shortauthors{W. Xue, J.-S. Niu, H.-F. Xue,  \& S. Yin}

\begin{document}
\begin{CJK*}{UTF8}{gbsn}

\title{Pulsation Analysis of High-Amplitude $\delta$ Scuti Stars with TESS}

\correspondingauthor{Jia-Shu Niu}
\email{jsniu@sxu.edu.cn}

\author{Wangjunting Xue (薛王俊婷)}
\affil{Institute of Theoretical Physics, Shanxi University, Taiyuan 030006, China}

\author[0000-0001-5232-9500]{Jia-Shu Niu (牛家树)}
\affil{Institute of Theoretical Physics, Shanxi University, Taiyuan 030006, China}
\affil{State Key Laboratory of Quantum Optics and Quantum Optics Devices, Shanxi University, Taiyuan 030006, China}

\author[0000-0001-6027-4562]{Hui-Fang Xue (薛会芳)}
\affil{Department of Physics, Taiyuan Normal University, Jinzhong 030619, China}
\affil{Institute of Computational and Applied Physics, Taiyuan Normal University, Jinzhong 030619, China}

\author{Sijing Yin (银思静)}
\affil{College of Physics and Electronic Engineering, Shanxi University, Taiyuan 030006, China}


\begin{abstract}
In this work, the pulsation analysis is performed on 83 high-amplitude $\delta$ Scuti stars, which have been observed by the Transiting Exoplanet Survey Satellite (TESS). The results show that 49 of these HADS show single-mode pulsation, 27 of them show radial double-modes pulsation (in which 22 of them pulsate with the fundamental and first overtone modes and 5 of them pulsate with the first and second overtone modes), and 7 of them show radial triple-modes pulsation (3 of which are newly confirmed triple-mode HADS). The histogram of the fundamental periods and the ratios between the fundamental and first overtone periods show bimodal structures, which might be caused by the stellar evolution in this specific phase. Most of the radial triple-mode HADS have a fundamental amplitude of 41-54 mmag, and 50\% of them have similar amplitudes of the fundamental and first overtone pulsation modes. All these hints require further confirmation not only in observations with more HADS samples, but also in theoretical models with suitable treatments of stellar evolution and pulsation.
\end{abstract}

\section{Introduction}           

$\delta$ Scuti stars are a classical type of short-period pulsating variable stars, which have periods range from 15 minutes to 8 hours and the spectral classes A-F. In Hertzsprung-Russell diagram, they locate on the main sequence (MS), pre-MS\footnote{A detailed guide of asteroseismology for pre-MS stars can be referred to \citet{Zwintz2022}.} or post-MS evolutionary stage at the bottom of the classical Cepheid instability strip and are self-excited by the $\kappa$ mechanism due to the partial ionization of helium in the out layers \citep{Breger2000,Kallinger2008,Handler2009,Guenther2009,Uytterhoeven2011,Holdsworth2014,Steindl2022}. 
Most of them pulsate in radial and non-radial p-mode \citep{Zong2015}, and some of them also show non-radial g-mode in low-frequency region simultaneously (hybrid pulsators, see e.g., \citet{Breger1996,Bradley2015,Yang2021}).

High-amplitude $\delta$ Scuti stars (hereafter HADS) are a subclass of $\delta$ Scuti stars, which have relatively larger amplitudes ($\Delta V \geq 0_{\cdot}^{m}1$) and slower rotations ($v \sin i \le 30\ \mathrm{km/s}$) in most cases. 
However, as the accumulation of the HADS samples, these classical criteria become unclear now (see e.g., \citet{Balona2012}). 
Most of HADS pulsate in single or double radial pulsation modes \citep{Niu2013,Alton2019,Bowman2021,Daszynska2022,Alton2022a,Alton2022b,Alton2022c}, and some of them have three radial pulsation modes \citep{Wils2008,Niu2022} or even some non-radial pulsation modes \citep{Poretti2011,Xue2020}.

On the aspect of stellar evolution theory, the period variation of a single star has significant different values in different evolutionary stages.
As a result, the observed linear period variation rate could be an important criterion to determine the evolutionary stage of a star.
Based on the times of maximum light lasting for decades, some HADS can be considered as normal stars evolving into a special evolutionary stage, which can be precisely determined according to asteroseismology self-consistently (see, e.g., \citet{Niu2017,Xue2018,Xue2022}). These results show that HADS should be located in the post-MS evolutionary stage.
However, some works show that some HADS can also be located in the terminal-age main-sequence (TAMS) or even MS (see, e.g., \citet{Bowman2021,Lv2022aj,Sun2021,Yang2022}), although the observed period variations are inconsistent with (always significantly greater than) the theoretical model predictions in these works.

These years, a large number of HADS have been monitored by the Transiting Exoplanet Survey Satellite (\TESS \citep{Ricker2015}), whose continues photometric data provide us to study their pulsation properties statistically.

\section{Methods}

We collected 10731 HADS in the International Variable Star Index (VSX), which were then performed cross matching with the input catalog of \TESS. At last, we got 83 HADS (most of which had 2-min cadence flux measurements, except some triple-mode HADS) from MAST Portal\footnote{https://mast.stsci.edu/portal/Mashup/Clients/Mast/Portal.html} which were processed by the \TESS Science Processing Operations Center (SPOC; \citet{Jenkins2016}).
After converting the normalized fluxes to magnitudes by utilizing the \TESS magnitude and removing the long trends in each Sectors, we chose the light curves in one Sector (which has the smallest value of standard deviation among all the Sectors) of each of the 83 HADS to perform the pre-whitening process.

In the pre-whitening process, the software Period04 \citep{Period04} was used to perform the Fourier transformations of the light curves to search for the significant peaks in the frequency spectra from 0 to 150 $\cd$, until there were no significant peaks (S/N $\ge$ 5.6 \citep{Zong2018}). After removing the alias frequencies considering the resolutions and gaps of the data sets, we obtained the significant frequencies and their amplitudes for each of the HADS. Each of the significant frequencies of an HADS was then identified whether it was an independent frequency or a harmonics/combination of some independent frequencies. At last, we got all the independent frequencies (and their amplitudes) for each of the HADS.

These independent frequencies were identified belonging to the radial pulsation modes based on the following relations \citep{Stellingwerf1979}: 
\begin{equation}
\begin{aligned}
  \label{eq:radial_modes}
  0.756 \leq &P_1/P_0 \leq 0.787,\\
  0.611 \leq &P_2/P_0 \leq 0.632,\\ 
  0.500 \leq &P_3/P_0 \leq 0.525,
\end{aligned}
\end{equation}
where $P_{0}$, $P_{1}$, $P_{2}$, and $P_{3}$ are considered as the periods of the fundamental, first overtone, second overtone, and the third overtone pulsation modes respectively.
 In this work, we would strictly follow the relations in Eq. (\ref{eq:radial_modes}) to perform the identification. Moreover, the independent frequencies which did not follow the relations in Eq. (\ref{eq:radial_modes}) were considered belonging to the non-radial pulsation modes, which we did not focus in this work.
For convenience, the fundamental mode, first overtone mode and second overtone mode are abbreviated as F, 1O, and 2O, respectively.\footnote{We did not find quadruple-mode HADS in this work.}

\section{Results}
In 83 HADS, we find 49 single-mode HADS, 27 radial double-mode HADS (22 with F and 1O pulsation and 5 with 1O and 2O pulsation), and 7 radial triple-mode HADS.

\subsection{Single-mode HADS}
In Table \ref{tab:single_mode}, we list the single-mode HADS and their periods and amplitudes. 
In the following statistical analysis, all these single-mode HADS are assumed to be pulsating in their fundamental modes, which might not be correct in all the cases and needs further research based on more information of the stars\footnote{An interesting work \citep{Pietrukowicz2020} finds that the shapes of the light curves are different between the single-mode fundamental and first overtone $\delta$ Scuti stars. But this needs a further check and requires clear quantitative criteria for application.}.

\begin{table*}[!htbp]
\begin{center}
  \caption{Periods and amplitudes of the single-mode HADS. }
  \label{tab:single_mode}
\resizebox{0.6\textwidth}{!}{
 \begin{tabular}{cccc}
   \hline
\hline   
   ID                & $P_0$(days) & $A_0$(mmag) &\TESS Sectors\\
\hline
 AD CMi & 0.12297 & 93.4 & 34$^{*}$ \\ 
AN Lyn & 0.09827 & 53.5 & 21$^{*}$\\ 
ASAS J065937-0047.0 & 0.10882 & 78.45 &  33$^{*}$\\ 
ASAS J015307-5056.5 & 0.091 & 87.02 &  2$^{*}$, 3, 29, 30 \\ 
BL Cam & 0.0391 & 98.76 &  19$^{*}$\\ 
BM For & 0.05411 & 74.56 &  4, 31$^{*}$ \\ 
BO Lyn & 0.09335 & 63.88 & 21$^{*}$, 47 \\ 
BS Aqr & 0.1978 & 119.28 & 2$^{*}$, 29, 42  \\ 
CC And & 0.1249 & 47.56 &  17$^{*}$ \\ 
CY Aqr & 0.06104 & 194.25 & 42$^{*}$ \\ 
DX Cet & 0.10396 & 56.69 & 31$^{*}$, 42, 43\\ 
GP And & 0.07868 & 160.33 &  17$^{*}$ \\ 
GSC 01951-01755 & 0.1207 & 165.6 &  45, 46$^{*}$ \\ 
GW Uma & 0.20317 & 121.15 & 21$^{*}$, 48 \\ 
KU Cen & 0.08 & 138.34 & 37$^{*}$\\ 
PT Com & 0.08211 & 86.44 & 22$^{*}$, 49  \\ 
RS Gru & 0.14701 & 154.7 & 1, 28$^{*}$\\ 
RY Lep & 0.22515 & 112.61 &  6, 32$^{*}$, 33 \\ 
SZ Lyn & 0.12053 & 145.8 & 20$^{*}$, 47 \\ 
TYC 3224-2602-1 & 0.08866 & 67.25 &  16$^{*}$ \\ 
V0337 Ori & 0.20129 & 127.27 & 43$^{*}$, 44, 45 \\ 
V0358 Mus & 0.12177 & 82.45 & 11$^{*}$, 12 \\ 
V0367 Cam & 0.12161 & 78.44 &  19$^{*}$ \\
V0398 Uma & 0.09134 & 122.29 &  15, 21, 22, 41$^{*}$, 48 \\ 
V0411 Sge & 0.13669 & 80.41 & 40$^{*}$ \\ 
V0451 Dra & 0.05581 & 169.4 &  14, 20, 21, 40$^{*}$, 41, 47, 48, 53 \\ 
V0467 Dra & 0.19768 & 112.31 &  15$^{*}$, 16, 23, 49, 50 \\ 
V0474 Mon & 0.13616 & 60.6 &  6$^{*}$, 33 \\ 
V0547 Lac & 0.09525 & 96.83 & 16$^{*}$\\ 
V0554 Vel & 0.10242 & 93.58 &  10, 36$^{*}$, 37\\ 
V0572 Cam & 0.08634 & 117.23 & 14$^{*}$, 20, 21, 40, 41, 53 \\ 
V0575 Lyr & 0.14555 & 57.46 & 40$^{*}$ \\ 
V0593 Lyr & 0.10215 & 181.1 &14, 40$^{*}$, 53, 54  \\ 
V0673 Hya & 0.10806 & 116.7 & 9$^{*}$, 36 \\ 
V0973 Cep & 0.08063 & 118.05 & 16-18, 19$^{*}$, 24, 49, 52 \\ 
V0974 Oph & 0.19102 & 124 &  12$^{*}$, 39\\ 
V1051 Ara & 0.11337 & 154.15 & 12, 39$^{*}$\\ 
V1307 Sco & 0.11703 & 167.86 &12, 39$^{*}$ \\ 
V1338 Cen & 0.13093 & 127.67 & 11, 38$^{*}$\\ 
V1421 Cen & 0.11417 & 131.93 & 37$^{*}$\\ 
V1429 Cen & 0.16701 & 54.69 &11$^{*}$ \\ 
V1535 Her & 0.09812 & 62.82 & 40$^{*}$\\ 
V2367 Cyg & 0.17665 & 129.02 & 14$^{*}$, 15, 40, 41, 54, 55\\ 
V2455 Cyg & 0.09421 & 137.28 &15, 16$^{*}$ \\ 
V5505 Sgr & 0.0844 & 136.27 &13$^{*}$, 39  \\ 
V6544 Sgr & 0.0632 & 146.7 & 13$^{*}$, 27\\ 
XX Cyg & 0.13486 & 227.83 & 14$^{*}$, 15-17, 41, 54-57\\ 
YZ Boo & 0.10409 & 119.91 & 24$^{*}$, 50, 51 \\ 
ZZ Mic & 0.06718 & 123.01 & 1, 27$^{*}$\\ 
\hline  
\hline
\end{tabular}
}
\end{center}
\footnotesize{Note: $^{*}$ denotes the Sector used in this work.}
\end{table*}

\subsection{Double-mode HADS}

In Table \ref{tab:double_mode}, we list the 27 radial double-mode HADS and their periods and amplitudes, together with the periods and amplitudes ratios of the double-modes.
What is of interest is that there are 5 HADS pulsating with the 1O and 2O modes (V0488 Gem is a newly confirmed such stars) and 5 HADS have larger amplitudes of 1O than F mode, which are denoted in Table \ref{tab:double_mode}.
The origins of these differences should be related to the pulsation mode selection mechanisms and worth in-depth studies in the future.

\begin{table*}[!htbp]
	\begin{center}
		\caption{Periods and amplitudes of the radial double-mode HADS. The ratios of the periods and amplitudes are also listed.}
    \label{tab:double_mode}
			\resizebox{\textwidth}{!}{
		\begin{tabular}{cccccccccccc}
			\hline
      \hline
		ID      & $P_{0}$(days) & $P_{1}$(days)   & $P_{2}$(days) & $P_{1}$/$P_{0}$   & $P_{2}$/$P_{1}$     & $A_{0}$(mmag) & $A_{1}$(mmag)     & $A_{2}$(mmag)    & $A_{1}$/$A_{0}$& $A_{2}$/$A_{1}$& \TESS Sectors\\
        \hline
		AI Vel   & 0.11158  & 0.08621    & -  & 0.77264 & -    & 93.61    & 65.79     & -                         & 0.70& -& 34$^{*}$, 35 \\                      
		BE Lyn              & 0.09588  & 0.07450                     & -        & 0.77705 & -                               & 114.65   & 2.59                       & -                         & 0.02  & -  & 21$^{*}$                      \\
		FP Cir              & 0.12675  & 0.09766                     & -        & 0.77049 & -                               & 78.34    & 2.84                       & -                         & 0.04  & -    & 12, 38$^{*}$                   \\
		GSC 04257-00471     & 0.17379  & 0.13307                     & -        & 0.76571 & -                               & 102.24   & 52.71                      & -                         & 0.52  & -                 & 16, 17, 18, 24$^{*}$       \\
		GSC 2583-00504      & 0.05172  & 0.03999                     & -        & 0.77321 & -                               & 54.61    & 18.31                      & -                         & 0.34  & -   & 24$^{*}$, 51, 52                  \\
      GSC 03949-00811$^{b}$     & 0.16977  & 0.13007                     & -        & 0.76615 & -                               & 6.21     & 8.27                       & -                         & 1.33  & -            & 16, 17, 41$^{*}$, 55           \\
      NSV 148000$^{b}$          & 0.15783  & 0.12207                     & -        & 0.77342 & -                               & 54.84    & 76.57                      & -                         & 1.40  & -     & 1, 2$^{*}$, 28                  \\
		NSVS 7293918        & 0.08854  & 0.06850                     & -        & 0.77369 & -                               & 118.69   & 8.54                       & -                         & 0.07  & -      & 20$^{*}$,44-47                  \\
		RV Ari              & 0.09312  & 0.07195                     & -        & 0.77263 & -                               & 135.90   & 39.67                      & -                         & 0.29  & -  & 42, 43$^{*}$                      \\
		SX Phe              & 0.05496  & 0.04277                     & -        & 0.77821 & -                               & 138.18   & 31.57                      & -                         & 0.23  & -    & 2$^{*}$, 29                    \\
		V0363 Tra           & 0.13856  & 0.10490                     & -        & 0.75710 & -                               & 140.34   & 3.06                       & -                         & 0.02  & -    & 12, 39$^{*}$                  \\
		V0388 Tel           & 0.14908  & 0.11277                     & -        & 0.75649 & -                               & 56.16    & 36.95                      & -                         & 0.66  & -     & 13$^{*}$, 27                   \\
      V0403 Gem$^{b}$           & 0.15339  & 0.11771                     & -        & 0.76735 & -                               & 42.42    & 43.94                      & -                         & 1.04  & -  & 43$^{*}$, 44, 45                   \\
		V0488 Gem$^{a}$           & -        & 0.09325                     & 0.07492  & -       & 0.80342                         & -        & 129.53                     & 0.80                      & -     & 0.01   & 33$^{*}$                  \\
      V0542 Cam$^{b}$           & 0.17477  & 0.13400                     & -        & 0.76675 & -                               & 71.09    & 84.71                      & -                         & 1.19  & -    & 19$^{*}$                \\
      V0703 Sco$^{b}$           & 0.14998  & 0.11521                     & -        & 0.76820 & -                               & 49.59    & 73.41                      & -                         & 1.48  & -    & 39$^{*}$              \\
		V0733 Pup           & 0.22876  & 0.17424                     & -        & 0.76167 & -                               & 107.69   & 47.62                      & -                         & 0.44  & -  & 7, 8, 34$^{*}$                   \\
		V0756 Cra           & 0.10719  & 0.08216                     & -        & 0.76651 & -                               & 104.07   & 30.02                      & -                         & 0.29  & -     & 13$^{*}$               \\
		V0798 Cyg$^{a}$           & -        & 0.19478                     & 0.15592  & -       & 0.80052                         & -        & 144.60                     & 15.60                     & -     & 0.11  & 14, 40, 41$^{*}$, 54                \\
		V0899 Car           & 0.11079  & 0.08585                     & -        & 0.77486 & -                               & 99.22    & 31.13                      & -                         & 0.31  & -   & 10, 11, 37$^{*}$                  \\
		V1049 Ara           & 0.10310  & 0.08050                     & -        & 0.78079 & -                               & 118.82   & 8.97                       & -                         & 0.08  & -    & 39$^{*}$                \\
		V1392 Tau           & 0.07443  & 0.05790                     & -        & 0.77789 & -                               & 81.86    & 43.89                      & -                         & 0.54  & -   & 5$^{*}$              \\
		V1553 Sco$^{a}$  & -         & 0.18429  & 0.14704                    & -       & 0.79783                         & -        & 52.45                      & 25.87                     & -     & 0.49   & 12, 39$^{*}$            \\
		V1719 Cyg$^{a}$  & -        & 0.26726  & 0.21374                     & -       & 0.79975                         & -        & 96.27                      & 10.31                     & -     & 0.11  & 15$^{*}$, 16, 55                 \\
		V2855 Ori           & 0.05808  & 0.04483                     & -        & 0.77177 & -                               & 88.59    & 24.70                      & -                         & 0.28  & -   & 6$^{*}$, 33                  \\
		VX Hya              & 0.22341  & 0.17274                     & -        & 0.77322 & -                               & 92.00    & 74.71                      & -                         & 0.81  & -  & 8, 35$^{*}$                  \\
		VZ Cnc$^{a}$    & -        & 0.17837 & 0.14280                  & -       & 0.80058 & -        & 125.09 & 50.04 & -     & 0.40 & 7, 34, 44, 45, 46$^{*}$    \\
      \hline
      \hline
		\end{tabular}
	}
\end{center}
\footnotesize{Note: $^{a}$ denotes the HADS pulsating with 1O and 2O; $^{b}$ denotes the HADS whose $A_{1}/A_{0} > 1$; $^{*}$ denotes the Sector used in this work.}
\end{table*}

\subsection{Triple-mode HADS}

In Table \ref{tab:triple_mode}, we list the 7 radial triple-mode HADS and their periods and amplitudes, together with the ratios of the periods and amplitudes of different pulsation modes.
We also note that 3 HADS have larger amplitudes of 1O than F, which might be related to the same cases in double-mode HADS.

Up to now, only fifteen HADS are confirmed to pulsate in the first three radial modes (the fundamental mode plus the first and second overtone) in the Galaxy: V829 Aql \citep{V829Aql}, GSC 762-110 (DO CMi) \citep{GSC762110}, GSC 03144-595 \citep{GSC03144-595}, GSC 08928-01300 \citep{Yang2021}, V761 Peg \citep{V761Peg}, two cases in \citet{Khruslov2014} (V0803 Aur and V1647 Sco), four cases in the OGLE project (OGLE-GD-DSCT-0021, OGLE-GD-DSCT-0033, OGLE-GD-DSCT-0048, and OGLE-GD-DSCT-0049), and four cases in \citet{Khruslov2022} (GSC2.3 NBY9001324, GSC2.3 NAVT000282, GSC2.3 S5SR001526, and GSC2.3 S4NM025687). In addition, two triple-mode HADS are confirmed in the Large Magellanic Cloud (OGLE-LMC-DSCT-0927 and OGLE-LMC-DSCT-2345) \citep{Poleski2010}. These stars exhibit different pulsating properties comparing with other HADS, which would help us to understand the selection mechanism of pulsation modes deeply \citep{Niu2022}.

\begin{figure}[!htbp]
  \centering
  \includegraphics[width=0.7\textwidth, angle=0]{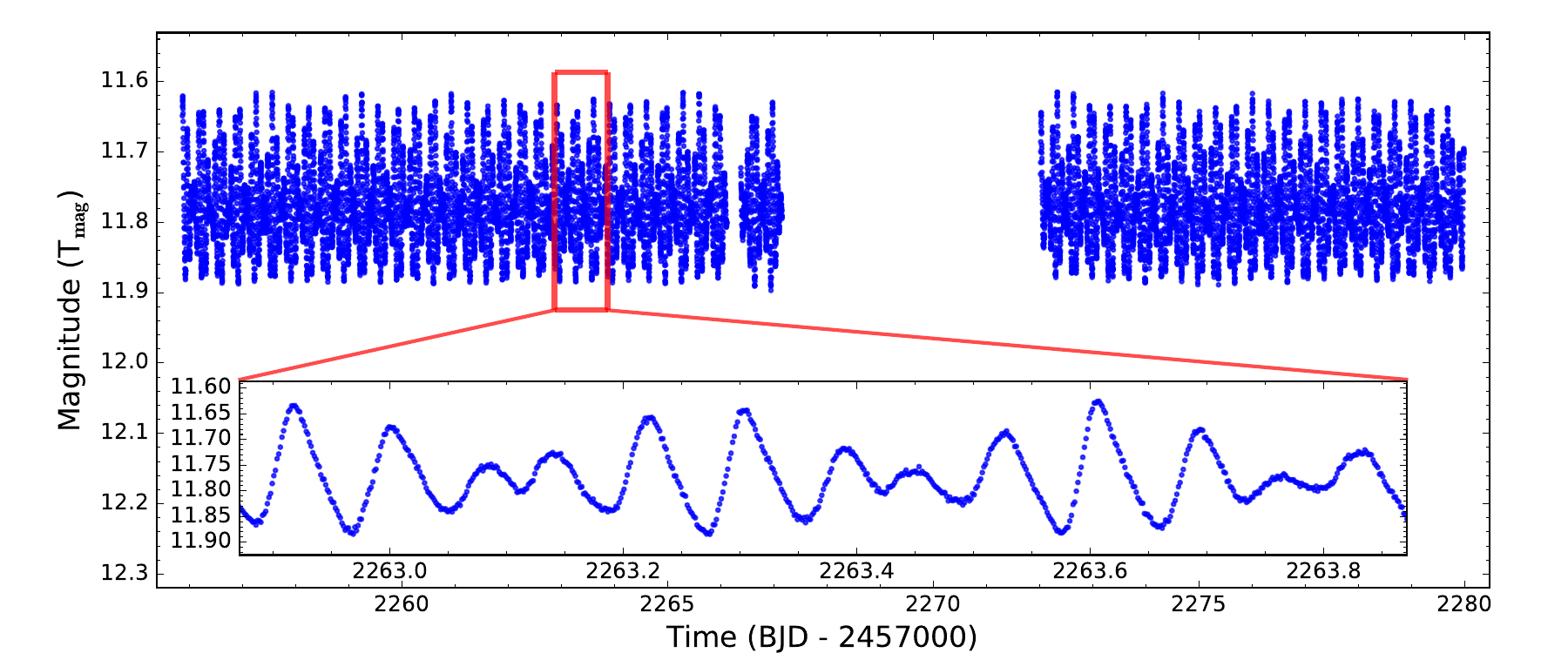}
  \includegraphics[width=0.7\textwidth, angle=0]{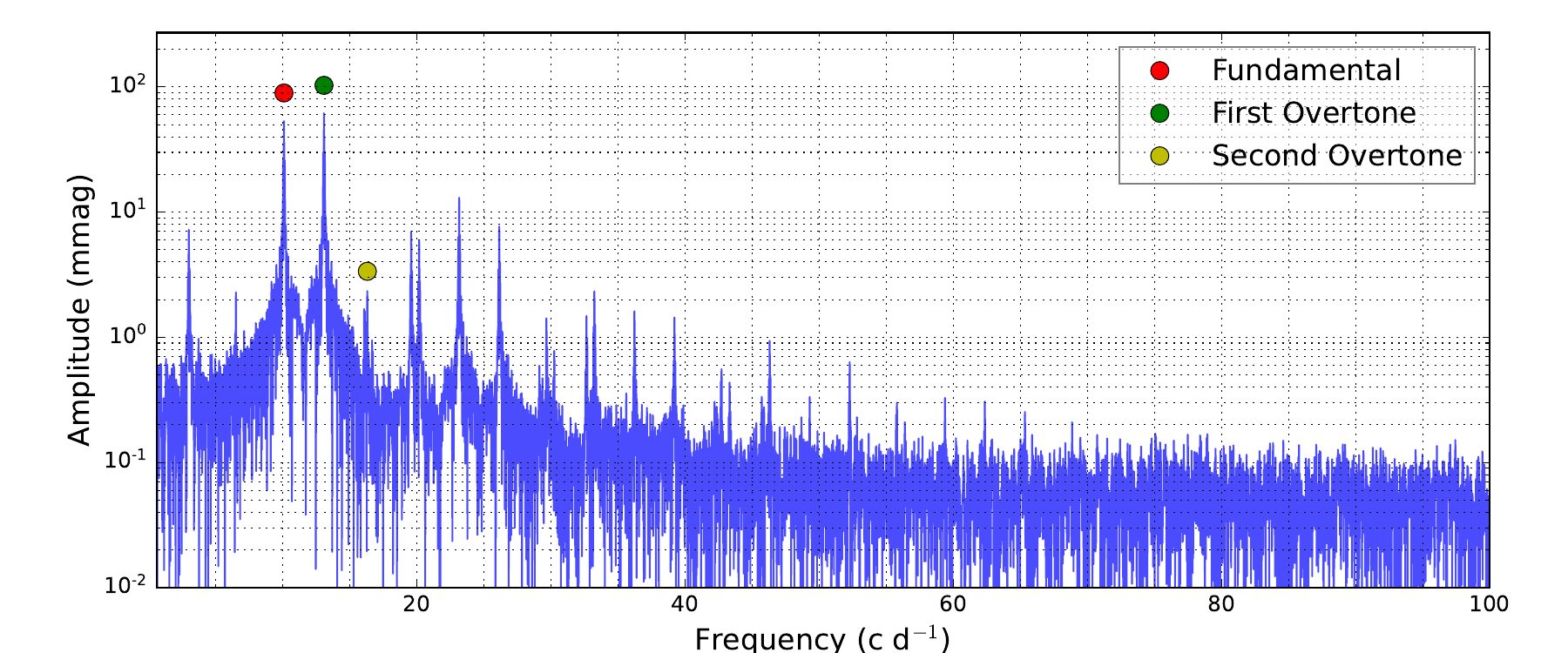}
  \caption{Light curves and frequency spectrum of ASAS J094303-1707.3 observed by \TESS.}
  \label{fig:lc_ASAS0917}
\end{figure}

\begin{figure}[!htbp]
  \centering
  \includegraphics[width=0.7\textwidth, angle=0]{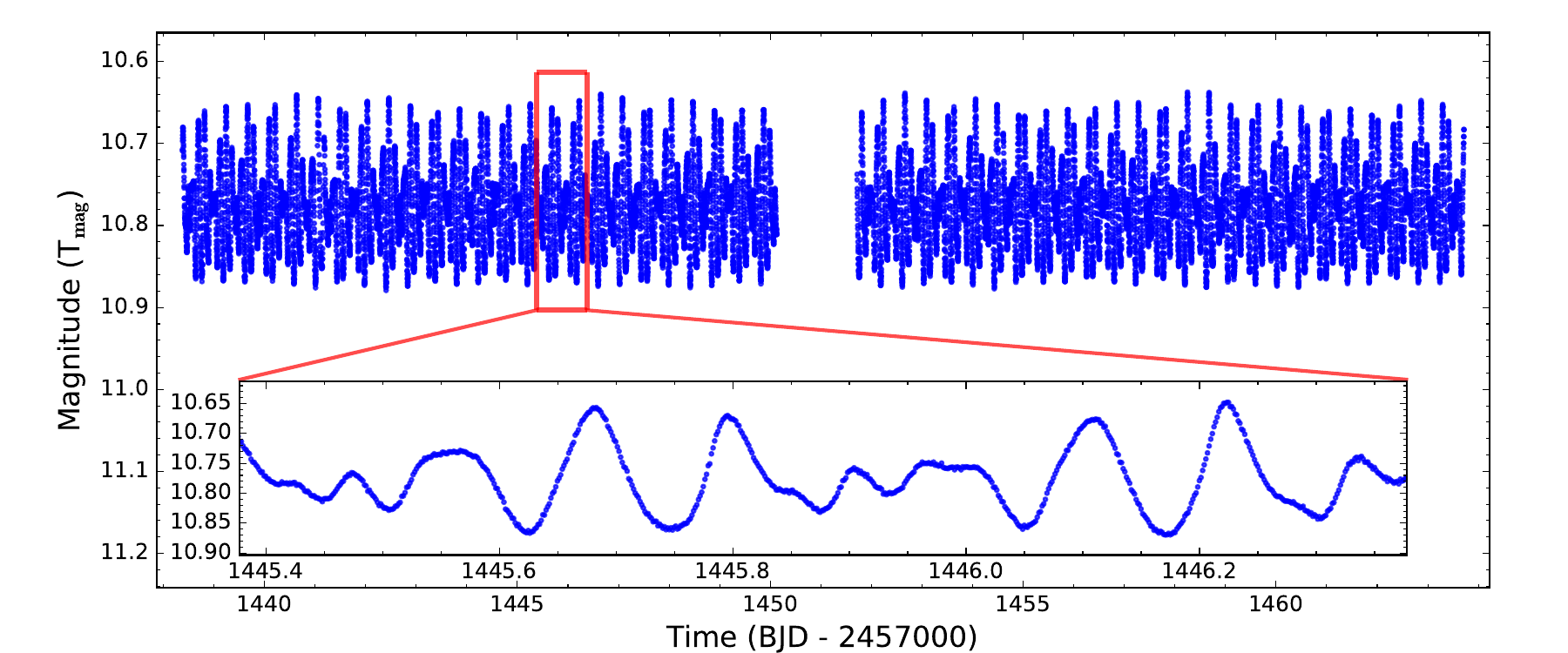}
  \includegraphics[width=0.7\textwidth, angle=0]{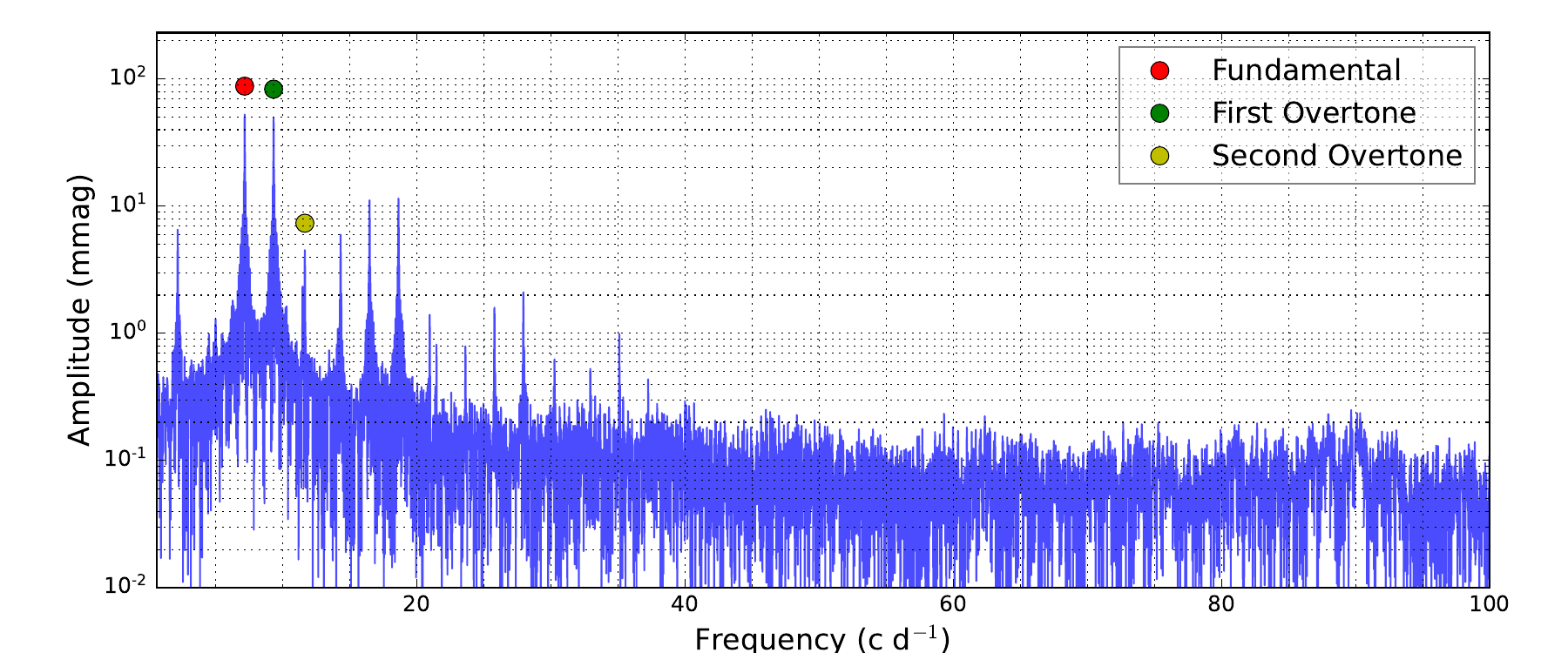}
  \caption{Light curves and frequency spectrum of V1384 Tau observed by \TESS.}
  \label{fig:lc_V1384Tau}
\end{figure}

\begin{figure}[!htbp]
  \centering
  \includegraphics[width=0.7\textwidth, angle=0]{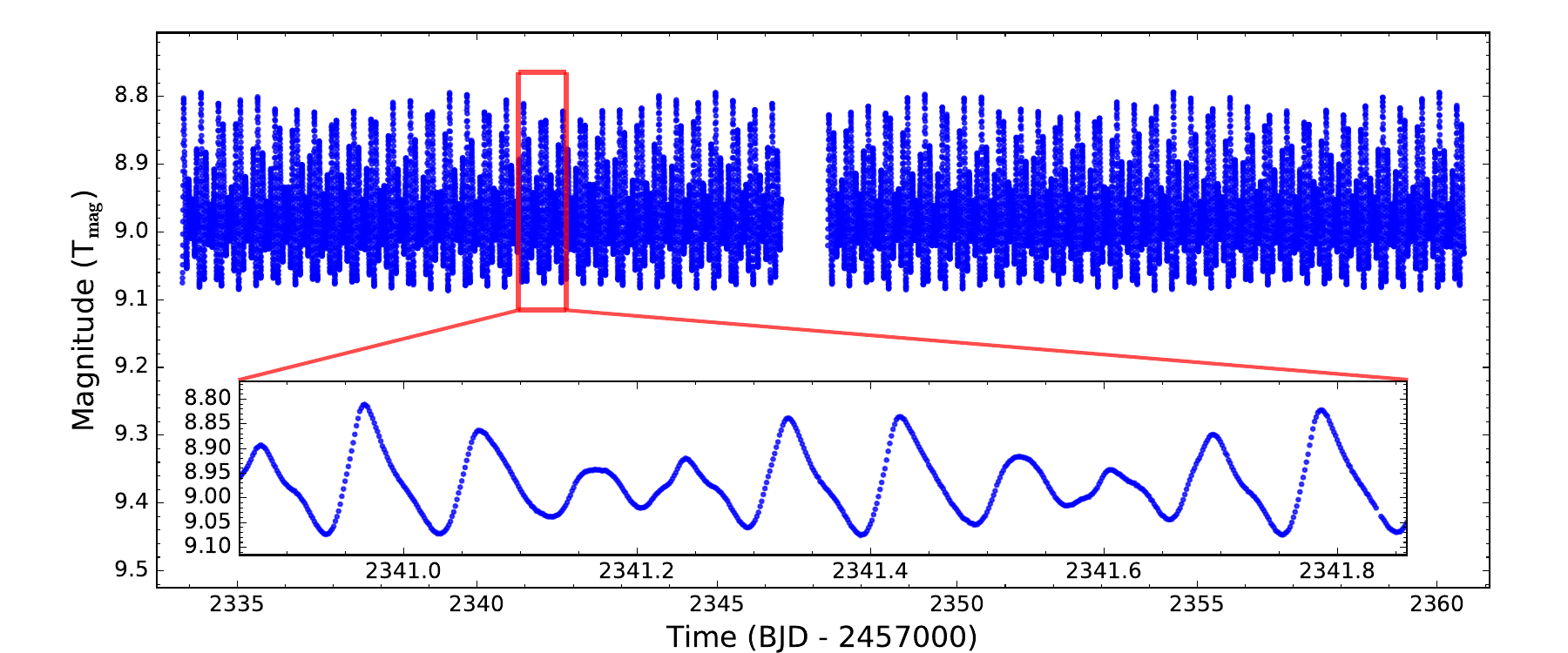}
  \includegraphics[width=0.7\textwidth, angle=0]{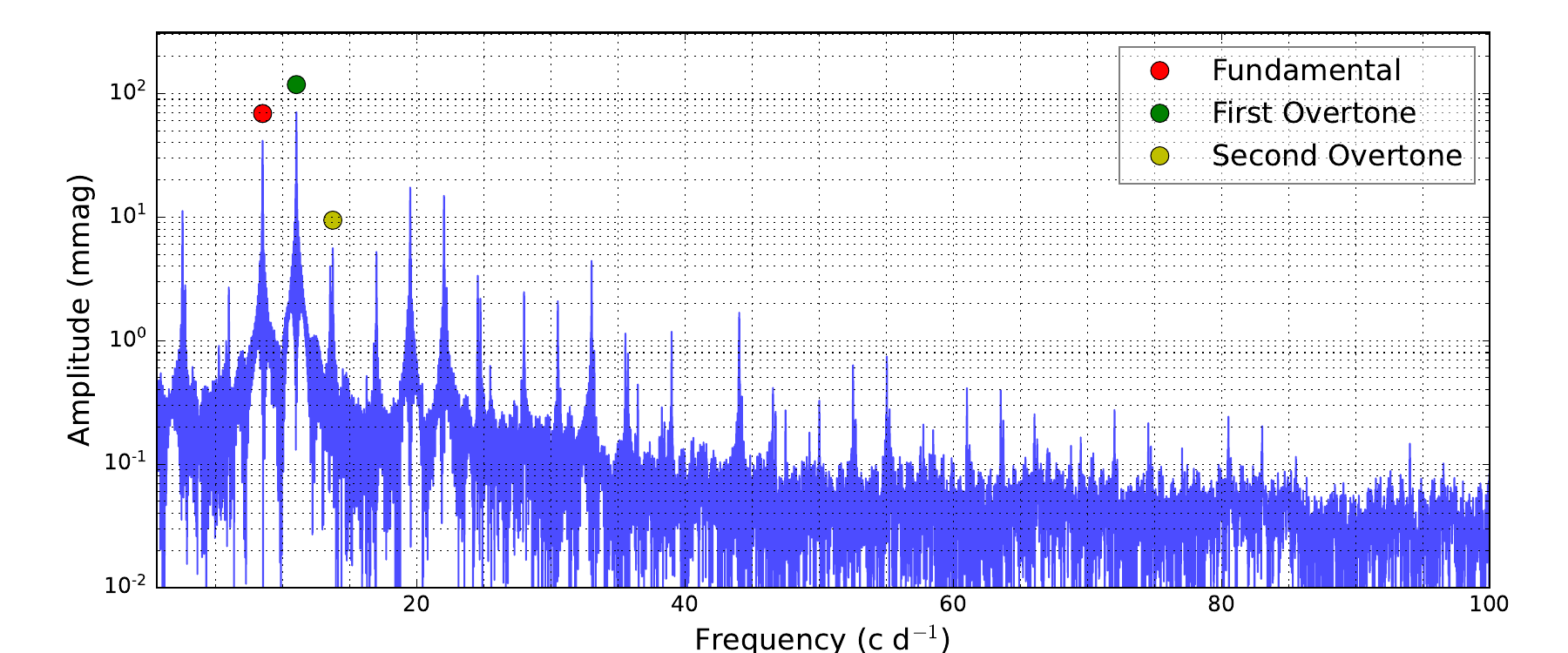}
  \caption{Light curves and frequency spectrum of V1393 Cen observed by \TESS.}
  \label{fig:lc_V1393Cen}
\end{figure}

In this work, three radial triple-mode HADS are newly confirmed: ASAS J094303-1707.3, V1384 Tau and V1393 Cen, and their light curves and frequency spectra are shown in Figure \ref{fig:lc_ASAS0917}, \ref{fig:lc_V1384Tau}, and \ref{fig:lc_V1393Cen}, respectively. The continuous photometric data provide us an opportunity to study the nature of them in the near future.

\begin{table*}[!htbp]
	\begin{center}
		\caption{Periods and amplitudes of the radial triple-mode HADS. The ratios of the periods and amplitudes are also listed.}
    \label{tab:triple_mode}
			\resizebox{\textwidth}{!}{
		\begin{tabular}{cccccccccccc}
			\hline
      \hline
			ID      & $P_{0}$(days) & $P_{1}$(days) & $P_{2}$(days) & $P_{1}/P_{0}$   & $P_{2}/P_{0}$   & $A_{0}$(mmag) & $A_{1}$(mmag) & $A_{2}$(mmag) & $A_{1}/A_{0}$ & $A_{2}/A_{0}$  &\TESS Sectors\\
      \hline
      ASAS J094303-1707.3$^{b}$    & 0.09918 & 0.07652 & 0.06134 & 0.77149  & 0.61852  & 53.54883 & 61.60368 & 2.01469 & 1.15042  & 0.03762& 8,35$^{*}$\\
      GSC 762-110$^{b}$  & 0.19450 & 0.14862 & 0.11909 & 0.76415  & 0.61228  & 48.98891 & 49.01288 &25.00426 & 1.00049  & 0.51041& 7(1800s), 33$^{*}$(600s)  \\
      GSC 03144-595& 0.20367 & 0.15548 & 0.12445 & 0.76338  & 0.61104  & 50.66162 & 49.35125 &11.25300 & 0.97413  & 0.22212& 14(1800s), 15$^{*}$(1800s)  \\
      GSC 08928-01300$^{c}$& 0.07573 & 0.05881 & 0.04750 & 0.77656  & 0.62720  & 76.66062 &  0.55071 & 0.91022 & 0.00718  & 0.01187&4,5,7-11,27,28,31,34,35,37,38$^{*}$ \\
      V0803 Aur    & 0.07106 & 0.05503 & 0.04439 & 0.77448  & 0.62473  & 49.48538 & 20.77060 & 4.49043 & 0.41973  & 0.09074& 43$^{*}$,44,45 \\
      V1384 Tau    & 0.13980 & 0.10739 & 0.08589 & 0.76820  & 0.61439  & 52.75469 & 49.93025 & 4.40794 & 0.94646  & 0.08356& 5$^{*}$ \\
      V1393 Cen$^{b}$    & 0.11778 & 0.09083 & 0.07285 & 0.77118  & 0.61850  & 41.30245 & 70.78809 & 5.65266 & 1.71390  & 0.13686 & 11,38$^{*}$ \\
      \hline
		\end{tabular}
	}
\end{center}
\footnotesize{Note: $^{b}$ denotes the HADS whose $A_{1}/A_{0} > 1$; $^{c}$ denotes the HADS is a $\delta$ Scuti and $\gamma$ Dor hybrid star; $^{*}$ denotes the Sector used in this work; GSC 762-110 and GSC 03144-595 do not have the data of 120s exposure, we used the 600s and 1800s data instead.}
\end{table*}

\section{Discussions and Conclusions}

Although the number of the samples is not that large, the statistical analysis of the pulsation properties of all the 83 HADS in Table \ref{tab:single_mode}, \ref{tab:double_mode}, and \ref{tab:triple_mode} could provide us some hints about the different origins of the three types.

\begin{figure}[!htbp]
  \centering
  \includegraphics[width=0.7\textwidth, angle=0]{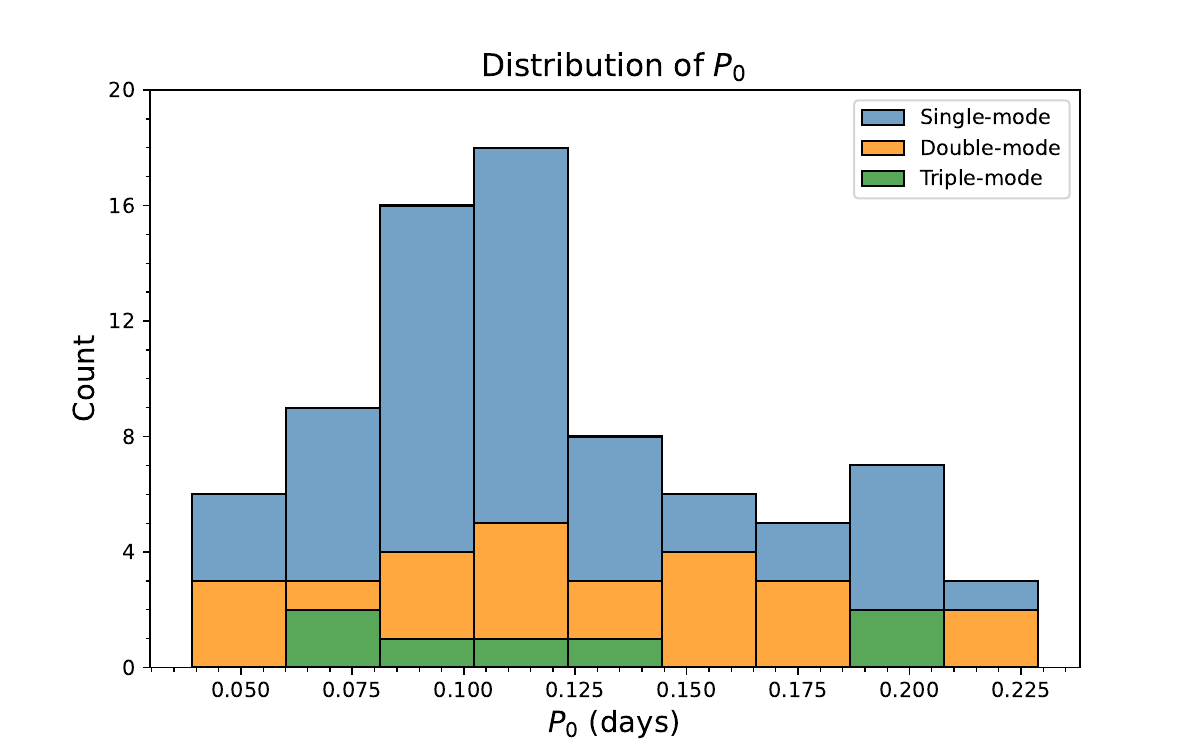}
  \caption{Distribution of $P_{0}$ of the 78 HADS.}
  \label{fig:dist_P0}
\end{figure}

Figure \ref{fig:dist_P0} shows the distribution of $P_{0}$ (period of the fundamental pulsation mode) of the 78 HADS (except the radial double-mode HADS pulsating with 1O and 2O), in which we can see that the most values of $P_{0}$ are concentrated in the range of 0.08 to 0.13 days. In detail, there is a small peak at about 0.2 days, which might be related to the two possible evolutionary stages of HADS (MS and post-MS). 

\begin{figure}[!htbp]
  \centering
  \includegraphics[width=0.7\textwidth, angle=0]{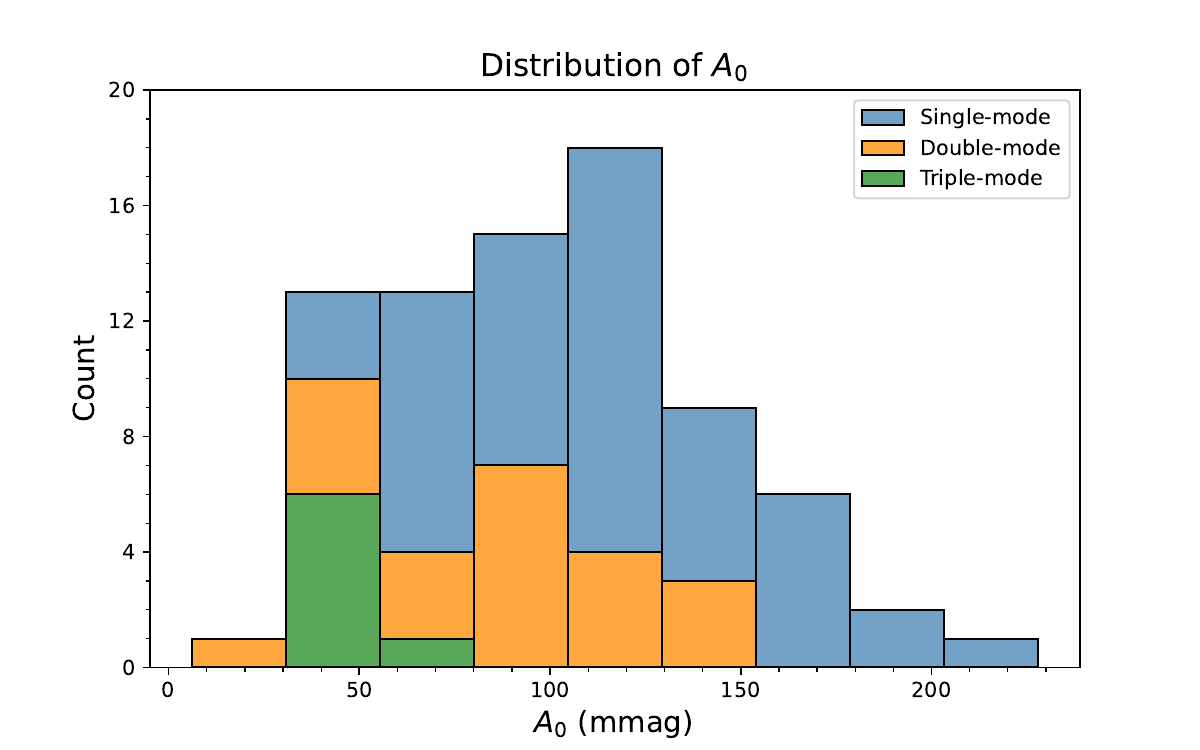}
  \caption{Distribution of $A_{0}$ of the 78 HADS.}
  \label{fig:dist_A0}
\end{figure}

Figure \ref{fig:dist_A0} shows the distribution of $A_{0}$ (amplitude of the fundamental pulsation mode) of the 78 HADS (except the radial double-mode HADS pulsating with 1O and 2O), in which we can see that the most values of $A_{0}$ are concentrated in the range of 30 to 130 mmag. The samples decrease slowly when $A_{0} > 130\ \mathrm{mmag}$ while drops sharply when  $A_{0} < 30\ \mathrm{mmag}$ (which might be caused by the selection criterion of the HADS samples). 
All the triple-mode HADS have the $A_{0}$ in the range of 30 to 80 mmag, if we consider the particularity of GSC 08928-01300 (a $\delta$ Scuti and $\gamma$ Dor hybrid star \citep{Yang2021}), all the other triple-mode HADS have the values of $A_{0}$ from about 41 to 54 mmag. The low-amplitude of these HADS can be explained by that more pulsation modes have to share the driving energy in the ionization zones, while the concentrated distribution of them might be a feature of this type of stars rather than a coincidence, which needs more samples to confirm.

\begin{figure}[!htbp]
  \centering
  \includegraphics[width=0.7\textwidth, angle=0]{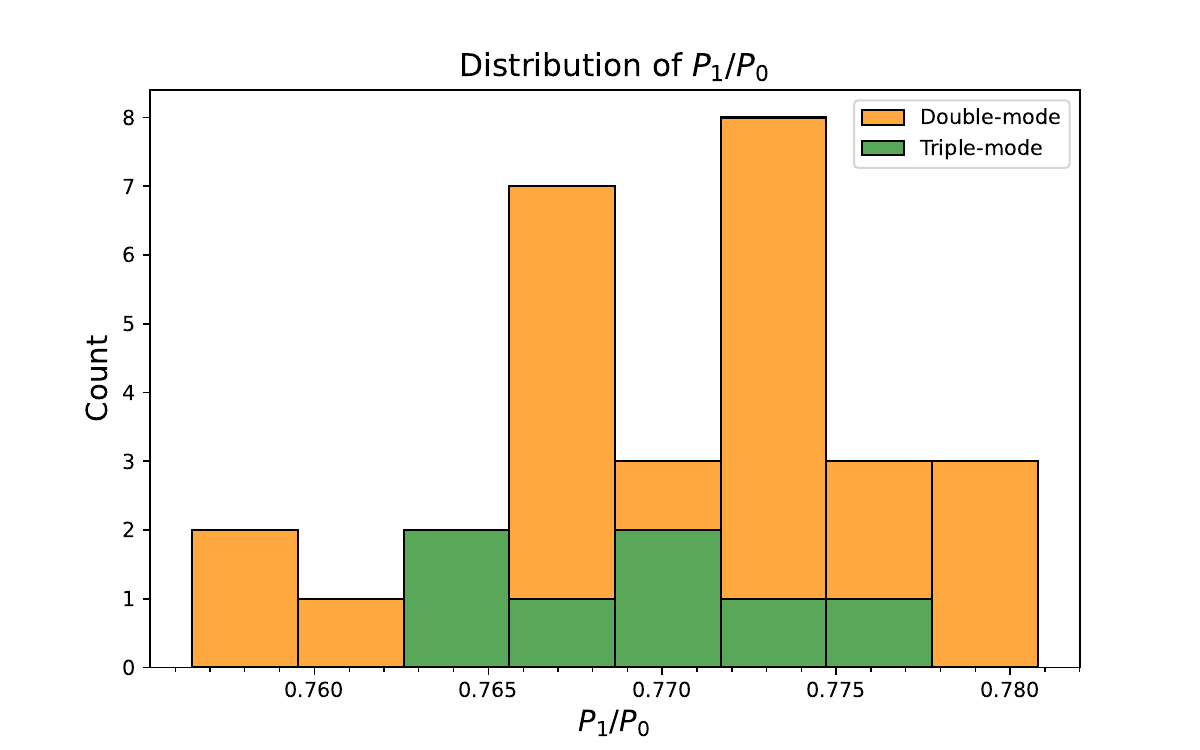}
  \caption{Distribution of $P_{1}/P_{0}$ of the 29 HADS.}
  \label{fig:dist_P1byP0}
\end{figure}

Figure \ref{fig:dist_P1byP0} shows the distribution of $P_{1}/P_{0}$ (periods ratio of the first overtone by the fundamental pulsation mode) of the 29 HADS (except the single-mode HADS and double-mode HADS pulsating with 1O and 2O), in which that of the double-mode shows an obvious bimodal structure, while that of the triple-mode show a relative smooth distribution from about 0.763 to 0.778.

\begin{figure}[!htbp]
  \centering
  \includegraphics[width=0.7\textwidth, angle=0]{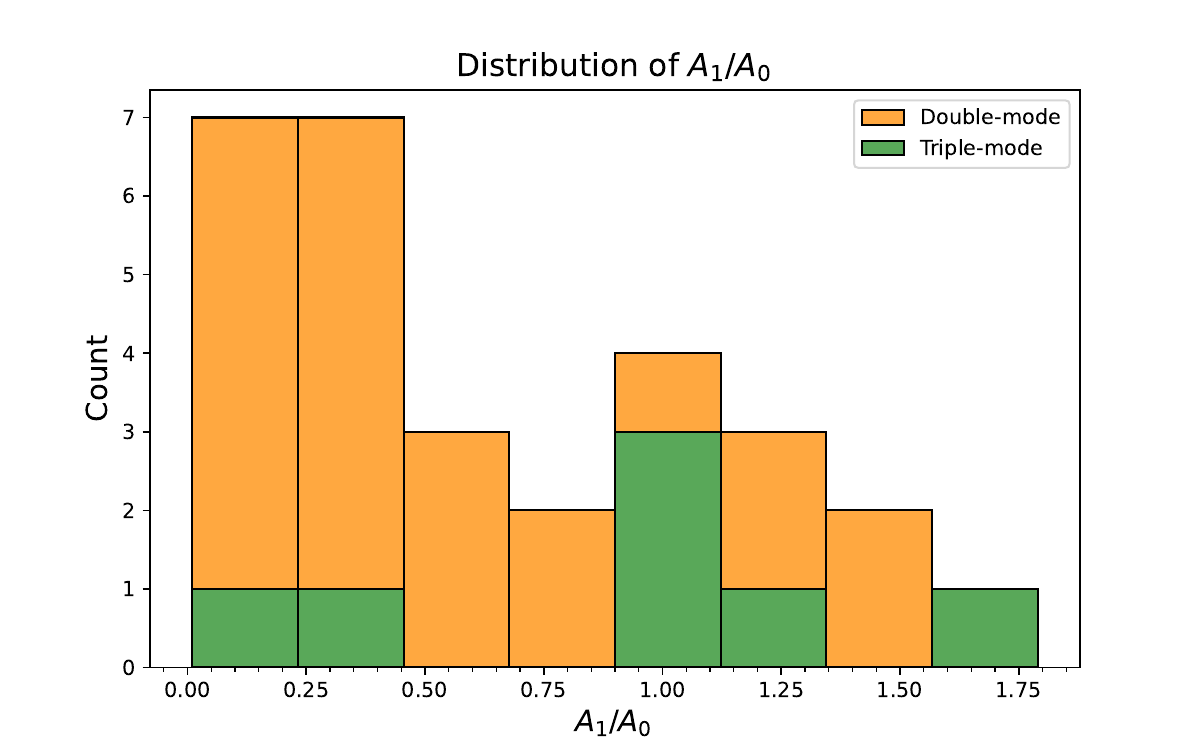}
  \caption{Distribution of $A_{1}/A_{0}$ of the 29 HADS.}
  \label{fig:dist_A1byA0}
\end{figure}

Figure \ref{fig:dist_A1byA0} shows the distribution of $A_{1}/A_{0}$ (amplitudes ratio of the first overtone by the fundamental pulsation mode) of the 29 HADS (except the single-mode HADS and double-mode HADS pulsating with 1O and 2O), in which we can see that the most values of $A_{1}/A_{0}$ of double-mode HADS are concentrated in the range of about 0 to 0.5. What is interesting is that, over 60\% of the triple-mode HADS have the values of $A_{1}/A_{0}$ around 1.0 (from about 0.8 to 1.2), which could be ascribed to the coupling between the fundamental and first overtone modes of these stars.
Such as the case in the triple-mode HADS KIC 6382916 (GSC 03144-595), the three pulsation modes are related by the resonance relationship $2f_1 + \Delta \omega = f_0 + f_2$ (where $\Delta \omega$ is the rotation frequency of the star), which couples the fundamental and first overtone modes together \citep{Niu2022}.

\begin{figure*}[!htbp]
  \centering
  \includegraphics[width=0.75\textwidth, angle=0]{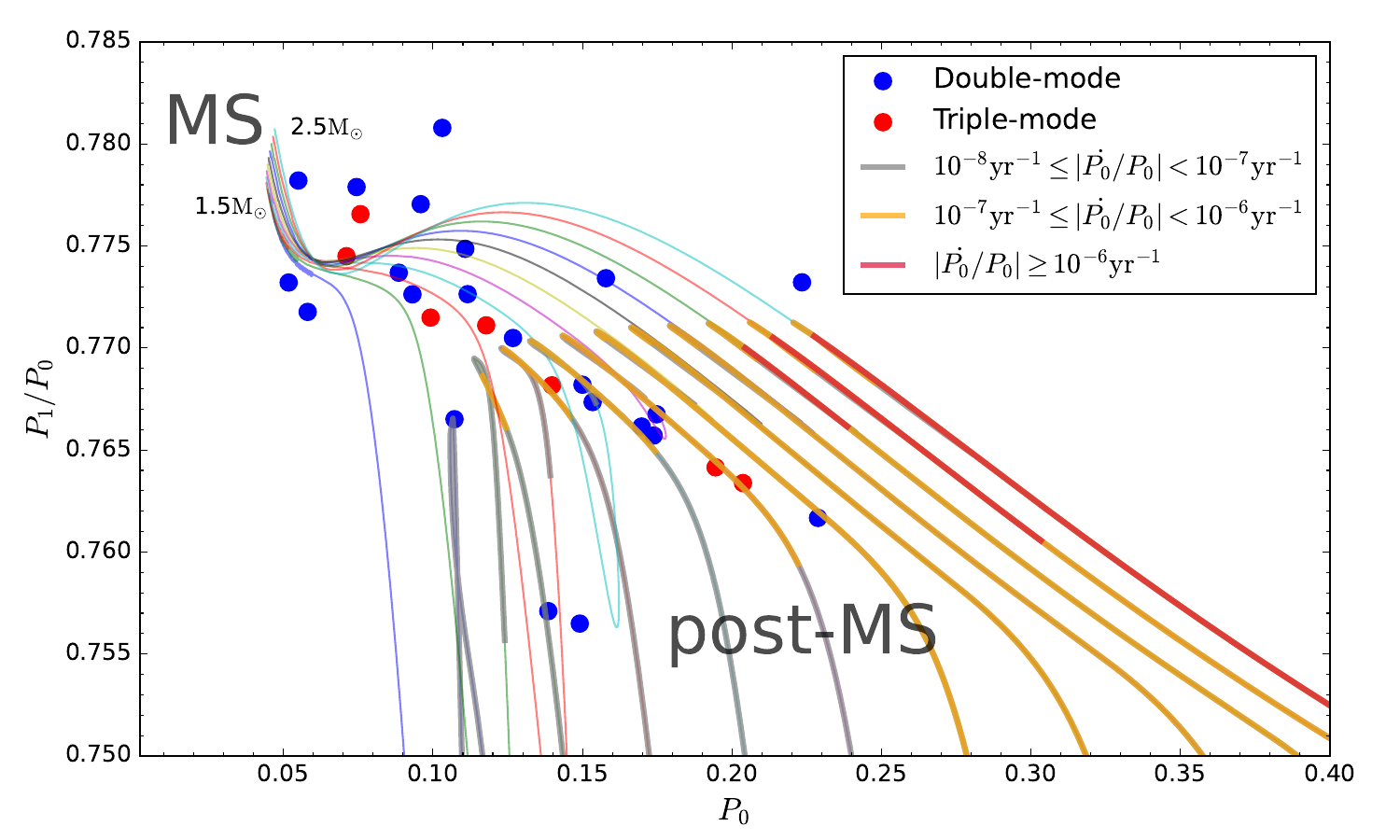}
  \includegraphics[width=0.75\textwidth, angle=0]{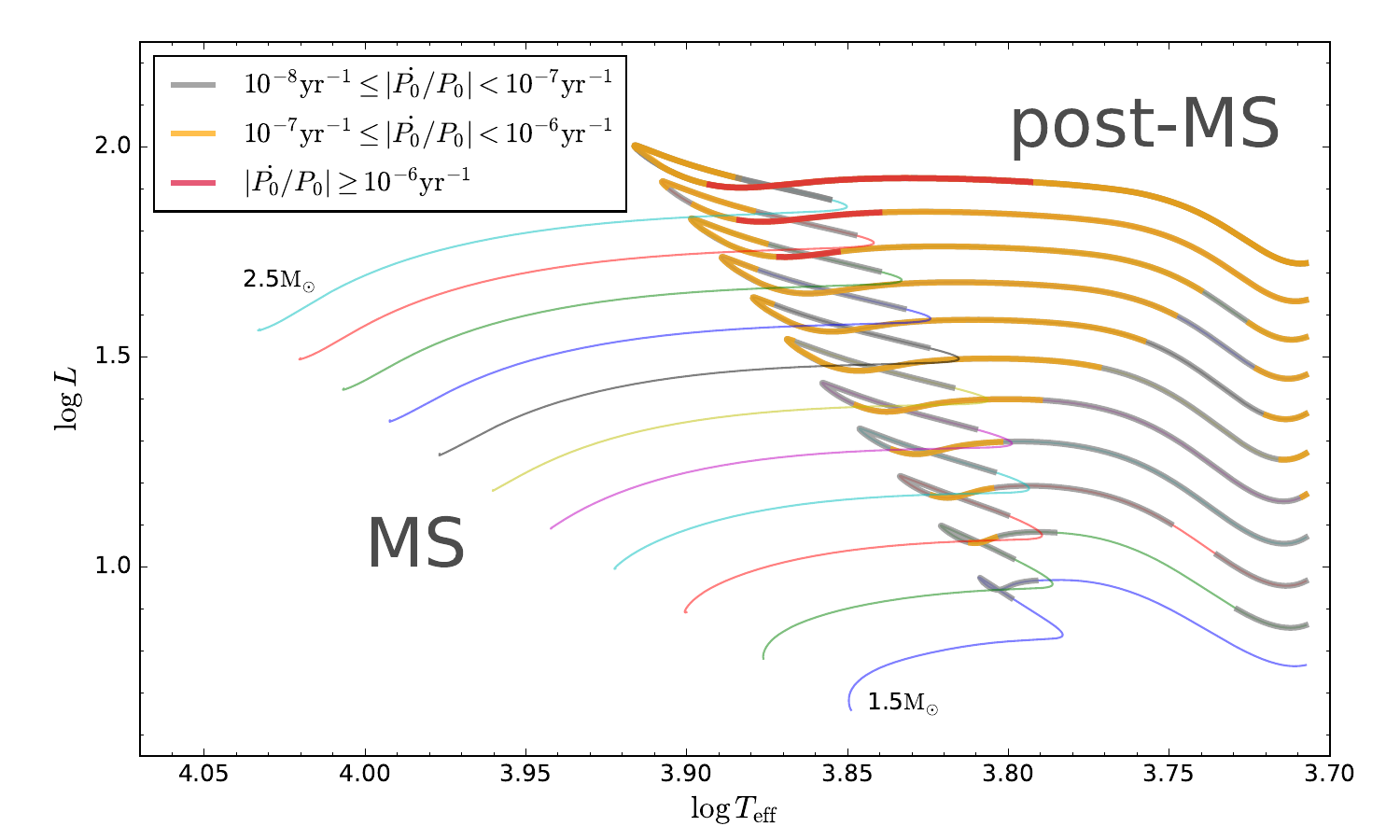}
  \caption{Petersen diagram (upper panel) of the 29 HADS and the relevant H-R diagram (lower panel) of the evolutionary tracks. The blue dots represent the double-mode HADS, and the red dots represent the triple-mode HADS. The colored thin lines represent the evolutionary tracks from MS to post-MS with masses from 1.5 to 2.5 $M_{\odot}$ with a step of 0.1 $M_{\odot}$, while the thick gray, orange, and red lines represent the evolutionary states whose $| \dot{P_0}/P_0 |$ fall into the range $10^{-8}\ \mathrm{yr}^{-1} \leq | \dot{P_0}/P_0 | < 10^{-7}\ \mathrm{yr}^{-1}$, $10^{-7}\ \mathrm{yr}^{-1} \leq | \dot{P_0}/P_0 | < 10^{-6}\ \mathrm{yr}^{-1}$, and $| \dot{P_0}/P_0 | \geq 10^{-6}\ \mathrm{yr}^{-1}$.}
  \label{fig:petersen}
\end{figure*}

In Figure \ref{fig:petersen}, we plot the Petersen diagram of the 29 HADS (who have fundamental and first overtone pulsation modes except the single-mode HADS and double-mode HADS pulsating with 1O and 2O), together with the evolutionary tracks of single stars (calculated by MESA \citep{Paxton2011,Paxton2013,Paxton2015,Paxton2018,Paxton2019} and GYRE \citep{Townsend2013,Townsend2018,Goldstein2020}, with the solar element abundances: $X=0.7438$, $Y=0.2423$, and $Z=0.0139$ \citep{Asplund2021}; the mixing-length parameter: $\alpha_{\mathrm{MLT}} = 1.89$ \citep{Niu2017}; the exponential scheme overshooting \citep{Herwig2000} whose parameter depends on the stellar mass $M$ (in solar masses): $f_{\mathrm{ov}} = (0.13 M - 0.098)/9.0$ \citep{Magic2010}) from MS to post-MS with masses from 1.5 to 2.5 $M_{\odot}$ with a step of 0.1 $M_{\odot}$.
As a comparison, the evolutionary tracks are also plotted in the Hertzsprung–Russell (H-R) diagram. The period change rate of the fundamental mode (which can be treated as a marker of the stellar evolutionary rate of the star) is also plotted in different regions: $10^{-8}\ \mathrm{yr}^{-1} \leq | \dot{P_0}/P_0 | < 10^{-7}\ \mathrm{yr}^{-1}$, $10^{-7}\ \mathrm{yr}^{-1} \leq | \dot{P_0}/P_0 | < 10^{-6}\ \mathrm{yr}^{-1}$, and $| \dot{P_0}/P_0 | \geq 10^{-6}\ \mathrm{yr}^{-1}$.

In each of the evolutionary tracks, the round evolutionary phase corresponds to the overall contraction phase after MS (i.e. the commonly known ``hook'' in the HR diagram), which is a quite rapid evolutionary phase in the range from MS to post-MS \citep{Aerts2010}. 
What is interesting is that these most rapidly evolving states in each of the evolutionary tracks start when $P_1/P_0 \sim 0.77$, which is clearly represented in the Petersen diagram of Figure \ref{fig:petersen}.
As a result, if we assume that the HADS are normal stars evolving in the MS to post-MS phase, the distribution of their $P_{1}/P_{0}$ would have a gap at about 0.77 and show a bimodal structure, which is consistent to the result shown in Figure \ref{fig:dist_P1byP0}. \footnote{Although the evolutionary tracks are just constructed based on the solar element abundances, the tracks with different element abundances have similar round evolutionary phases (even though they will have overall panning). If we combine all these tracks together, there still exists a most rapidly evolving state in the Petersen diagram around $P_{1}/P_{0} \sim 0.77$.} 
It gives us a hint that the pulsation properties of HADS might be related to their evolutionary phases, which should be tested and confirmed with more samples in the future.

\section*{Acknowledgments}
J.S.N. acknowledges support from the National Natural Science Foundation of China (NSFC) (No. 12005124 and No. 12147215). 
H.F.X. acknowledges support from the Scientific and Technological Innovation Programs of Higher Education Institutions in Shanxi (STIP) (No. 2020L0528) and the Applied Basic Research Programs of Natural Science Foundation of Shanxi Province (No. 202103021223320).


\label{lastpage}

\end{CJK*}
\end{document}